\documentstyle[12pt,here,epsfig]{article}
\begin{document}
{ \Large \bf
STUDY OF $K_{S}K_{L}$ COUPLED DECAYS AND $K_{L}$-Be 
INTERACTIONS WITH THE CMD-2 
DETECTOR AT VEPP-2M COLLIDER}
\begin{center}
R.R.Akhmetshin, G.A.Aksenov, E.V.Anashkin, V.M.Aulchenko,
\\
B.O.Baibusinov, V.S.Banzarov,  L.M.Barkov, S.E.Baru,  A.E.Bondar,
\\
D.V.Chernyak, V.V.Danilov, S.I.Eidelman, G.V.Fedotovich, N.I.Gabyshev,
\\
A.A.Grebeniuk, D.N.Grigoriev, P.M.Ivanov, B.I.Khazin, I.A.Koop, A.S.Kuzmin,
\\
I.B.Logashenko, P.A.Lukin, A.P.Lysenko, A.V.Maksimov, Yu.I.Merzlyakov, 
\\
I.N.Nesterenko, V.S.Okhapkin, E.A.Perevedentsev, A.A.Polunin, E.V.Popkov,
\\
V.I.Ptitzyn, T.A.Purlats, S.I.Redin, N.I.Root, A.A.Ruban, N.M.Ryskulov,
\\
Yu.M.Shatunov, A.E.Sher, M.A.Shubin, B.A.Shwartz, V.A.Sidorov,
\\
A.N.Skrinsky, V.P.Smakhtin, I.G.Snopkov, E.P.Solodov, A.I.Sukhanov,
\\
V.M.Titov, Yu.V.Yudin, V.G.Zavarzin
\\
  Budker Institute of Nuclear Physics, Novosibirsk, 630090, Russia
\\
\end{center}
\vspace {0.1cm}
\begin{center}
           D.H.Brown, J.P.Miller, B.L.Roberts, W.A.Worstell
\\
                   Boston University, Boston, MA 02215, USA
\\
\end{center}
\vspace{0.1cm}
\begin{center}
                J.A.Thompson, C.M.Valine
\\
                University of Pittsburgh, Pittsburgh, PA 15260, USA
\\
\end{center}
\vspace {0.1cm}
\begin{center}
            P.B.Cushman 
\\
            University of Minnesota, Minneapolis, Minnesota 55455, USA
\\
\end{center}
\vspace{0.1cm}
\begin{center}
             S.K.Dhawan, V.W.Hughes
\\
                    Yale University, New Haven, CT 06511, USA 
\end{center}
\vspace{.25cm}

%
\newpage
\begin{abstract}
\hspace*{\parindent}
The integrated luminosity $ \approx 4000 nb^{-1}$ of 
around $\phi$ meson mass ($ 5.0 \times 10^{6}$ of $\phi$'s)
has been collected with the CMD-2 detector at the VEPP-2M collider.
A latest analysis of the $K_{S}K_{L}$ coupled decays
based on 30\% of available data is presented in this paper.

The $K_{S}K_{L}$ pairs from $\phi$ decays were
reconstructed in the drift chamber when both kaons 
decayed into two charged particles.
From a sample of 1423 coupled decays a selection of candidates to
the CP violating $K_{L}\rightarrow\pi^+\pi^-$ decay was performed.
CP violating decays were not identified because of the domination of
events with a $K_{L}$ regenerating 
at the Be beam pipe into $K_{S}$ and a background from $K_{L}$ 
semileptonic decays.

The regeneration cross section of 110 MeV/c $K_{L}$'s  
was found to be $\sigma_{reg}^{Be}=53\pm17$ mb in agreement with theoretical
expectations. The angular distribution of $K_{S}$'s after regeneration  
and the total cross section of $K_{L}$ for Be have been measured.
\end{abstract}
\baselineskip=17pt
\section{Introduction}
\hspace*{\parindent}

      As was realized at the very early steps of the
$\phi$ meson studies at the colliding beam machines, $K_{S}K_{L}$ pairs
( $\approx 34\%$ of all $\phi$ decays) can be used as a 
source for studying CP and CPT violation. These suggestions, including
studies of quantum mechanical correlations, were discussed 
for experiments at the Novosibirsk  electron-positron
collider VEPP-2M \cite{bayer73}-\cite{vepp2m}.
The coupled decays of the $K_{S}K_{L}$  pairs will allow
demonstration of the quantum mechanical correlations
of the two particle decays (Einstein-Podolsky-Rosen paradox)\cite{epr}.

The $\phi$ resonance produced
in $e^{+}e^{-}$ collisions is also  a source of 
 low momentum neutral
and charged kaon pairs not available from other sources 
for studies of nuclear interactions. 
Mesons in each pair are produced with opposite and equal momenta and
a detector with good resolution and 
reconstruction efficiency allows one to use one reconstructed kaon as a tag for
another. The most natural way is to use a reconstructed decay 
of $K_{S} \rightarrow \pi^{+}\pi^{-}$ as a tag for $K_{L}$. 
In this case the momentum and direction of $K_{L}$ are completely
determined. 

The idea of creating an intensive source of $\phi$ mesons
has been discussed by  many   authors \cite{rosner,phigroup}.
 The  flux  of events   at   these   so-called   "$\phi$-factories"  now
 under  construction \cite{skrinsky91,vignola92} will make
feasible new
precise measurements of a possible direct component in  the  decay
$K_{L} \rightarrow  \pi^{+} \pi^{-}, \pi^{0} \pi^{0}$
($\epsilon'/\epsilon$),  as  well  as 
an observation of CP-violating three pion decays of the $K_{S}$.
Studies of the oscillations in the joint decay distributions
could provide information about
real and imaginary parts of
any CP-violating amplitude.

At  the  VEPP-2M collider
at  Novosibirsk, which could be considered as a pre $\phi$-factory,
we have been running with the CMD-2 detector  
preparing for experiments
at the $\phi$-factory which is under construction here.
Studies of an upgraded detector and accelerator
are in progress,  including an intermediate
$10^{32} cm^{-2}s^{-1}$ luminosity collider for investigating
the idea of the round beams, an important ingredient of the 
Novosibirsk $\phi$-factory project\cite{round,est90}.
\section{The CMD-2 Detector}
\hspace*{\parindent}
The  CMD-2  detector  has  been
described  in  more  detail  elsewhere \cite{CMD285,cmd2gen}.
The main systems of the detector are shown in Figure~1.

The CsI barrel calorimeter with a 6 x 6 x 15 $\mathrm cm^{3}$ crystal size 
is placed outside of
a 0.4 r.l. superconducting solenoid with a 1 Tesla 
magnetic field.  The endcap calorimeter is made of 2.5 x 2.5 x 15 
$\mathrm cm^{3}$
BGO crystals and has not been installed for the data presented here.

The drift chamber(DC) with a 30 cm outer radius
and a 44 cm length has about
 250 $\mu$m resolution transverse to the beam and  0.5 cm
longitudinally and is placed inside the solenoid.
The vertex reconstruction resolution for the neutral
kaon decays into charged particles is about 0.15 cm radially.
The muon range system consists of streamer tubes and has
1-3 cm spatial resolution.

 A 3.4 cm diameter vacuum beam pipe is made of Be with a 0.077 cm wall 
thickness and may be considered as a target for studies of the 
kaon nuclear interaction.

A data sample of the integrated
luminosity of 1500 $nb^{-1}$ has been analyzed around the $\phi$ 
corresponding to 
about $1.7 \times 10^{6}$ produced $\phi$'s.
 About 300 $nb^{-1}$ were used to
measure the $\phi$ meson parameters and its branching ratios into
four major decay modes \cite{PhLet95}.
 The largest part
of the integrated luminosity $ \approx 1200 ~nb^{-1}$
was used for the studies
of rare decay modes of $\phi$.
Some preliminary results were published in \cite{CMD2pre93}.

 These data were also intensively
used for studies of the detector performance and software development. 
 Several versions of the reconstruction program have been tested and 
a reconstruction efficiency of about 95-97\% per charged track has been
achieved with a momentum
uncertainty of 4$\%$ for 250 MeV/c  charged particles.
The energy resolution of 8$\%$ for photons in the CsI calorimeter
has been obtained.
 
The number of reconstructed 
$K_{S}K_{L}$ coupled decay events is now twice 
the sample in the preliminary
publication \cite{CMD2pre93}.
 The latest analysis of the $K_{S}K_{L}$ coupled decays 
is presented in this paper.
\begin{figure} [tbh]
  \begin{center}
   \begin{tabular}[t]{lr}
\mbox{\psfig{file=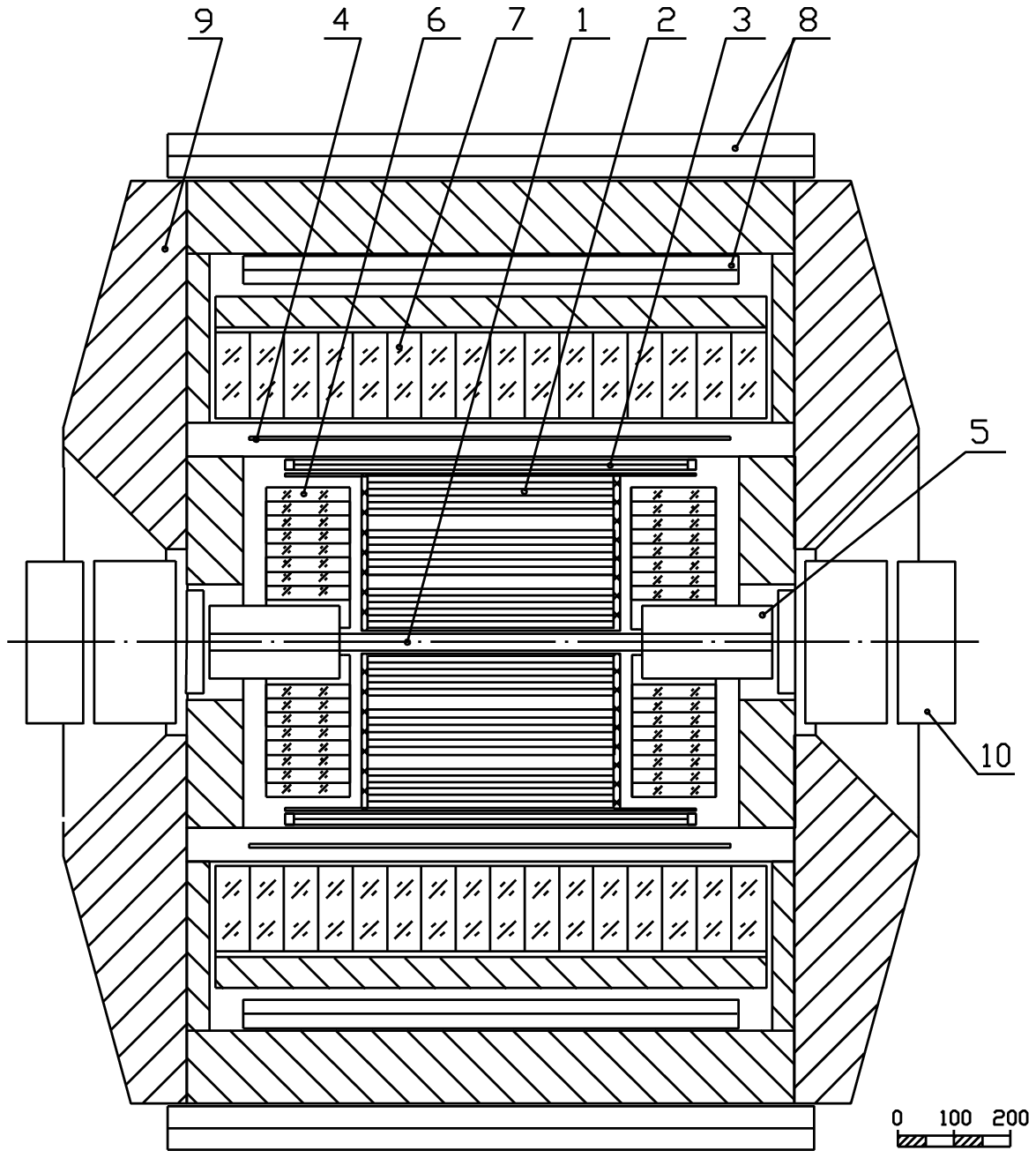,width=0.45\textwidth}} &
\mbox{\psfig{file=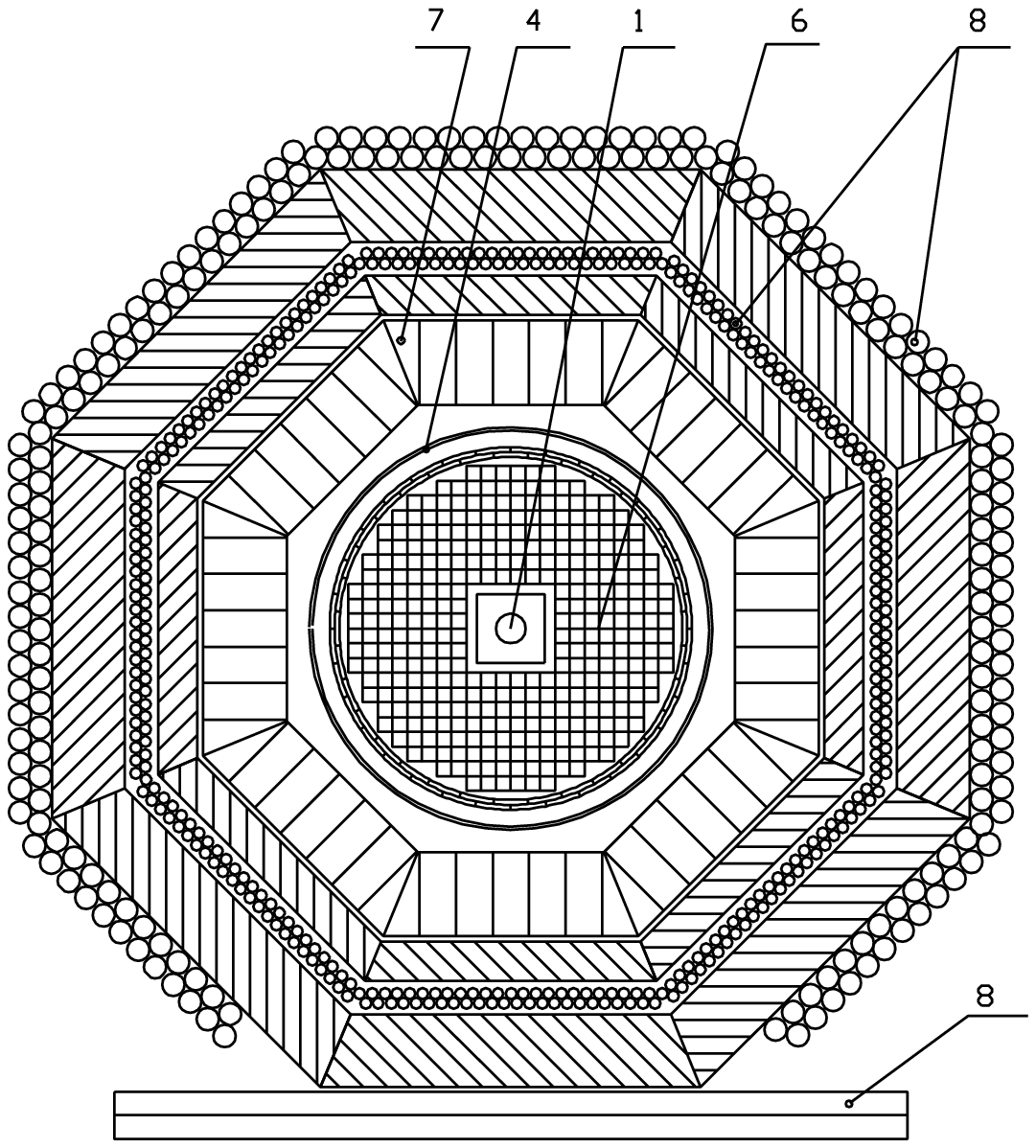,width=0.5\textwidth}}
   \end{tabular}
\caption{ Horizontal and vertical
cross sections of the CMD-2 detector. 1 - vacuum chamber;
2 - drift chamber; 3 - Z-chamber; 4 - main solenoid;
5 - compensating solenoid; 6 - BGO endcap calorimeter;
7 - CsI barrel calorimeter; 8 - muon range system; 9 - magnet yoke; 
10 - collider lenses.}
  \end{center}
\label{cmd2fig}
\end{figure}
\section{Selection of $K_{S}K_{L}$ coupled decays}
\begin{figure}[tbh]
\begin{center}
\begin{tabular}{cc}
\mbox{\psfig{file=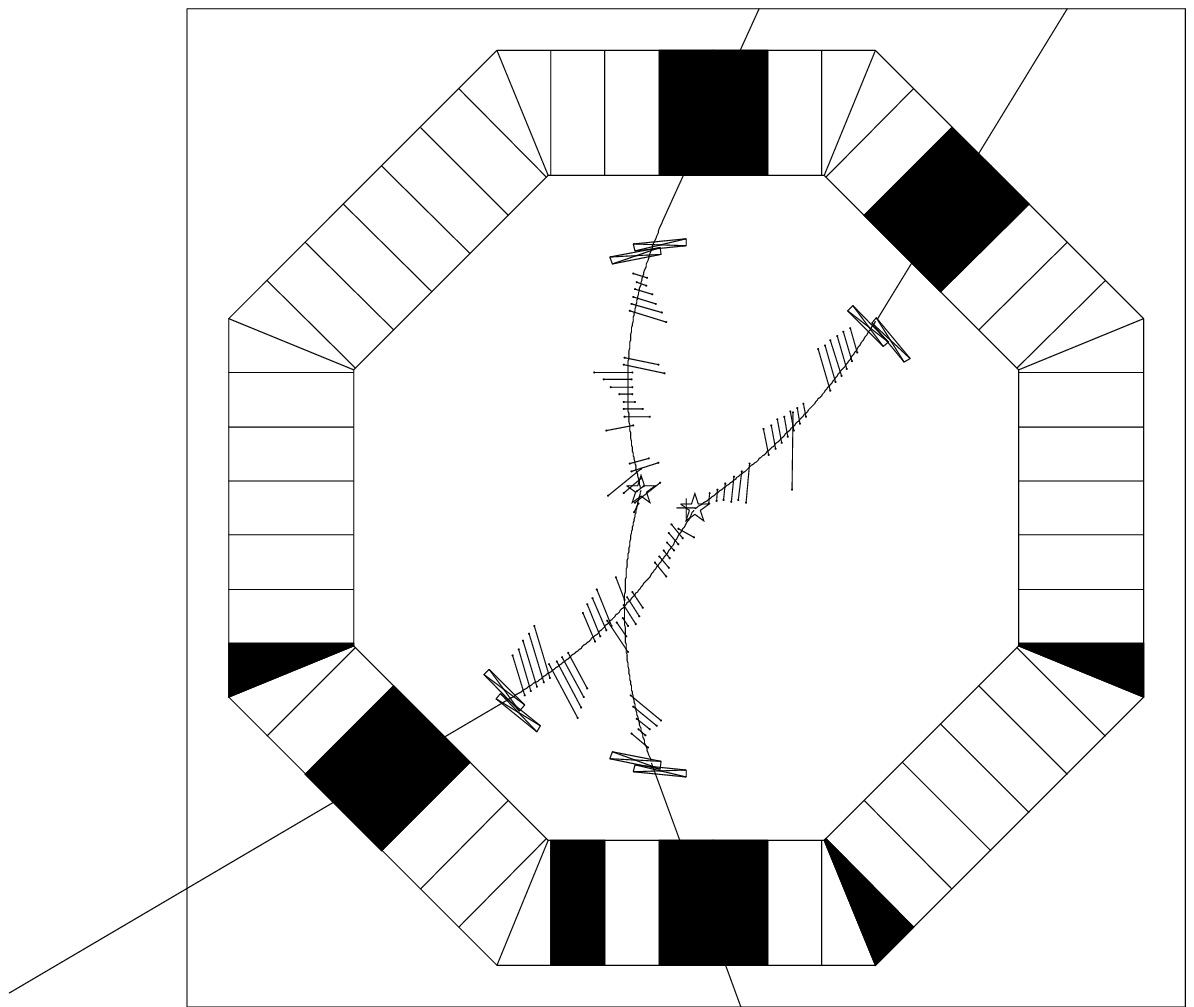,width=9cm}} &
\hspace{-3cm} \mbox{\psfig{file=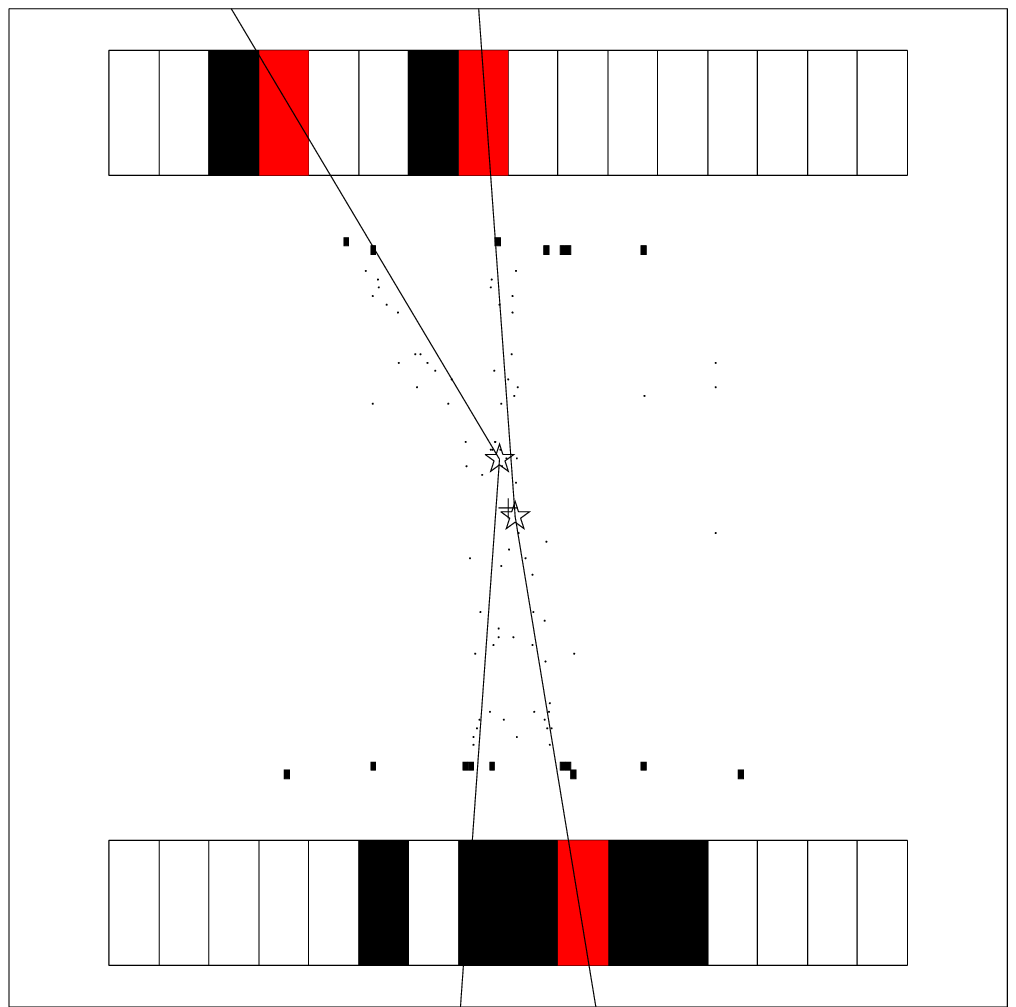,width=9cm}} 
\end{tabular}
\caption{$\phi\rightarrow K_{S}K_{L}$ event with a coupled decay.}
\label{kskldec.fig}
\end{center}
\end{figure}

\begin{figure}[H]
\begin{center}
\mbox{\psfig{figure=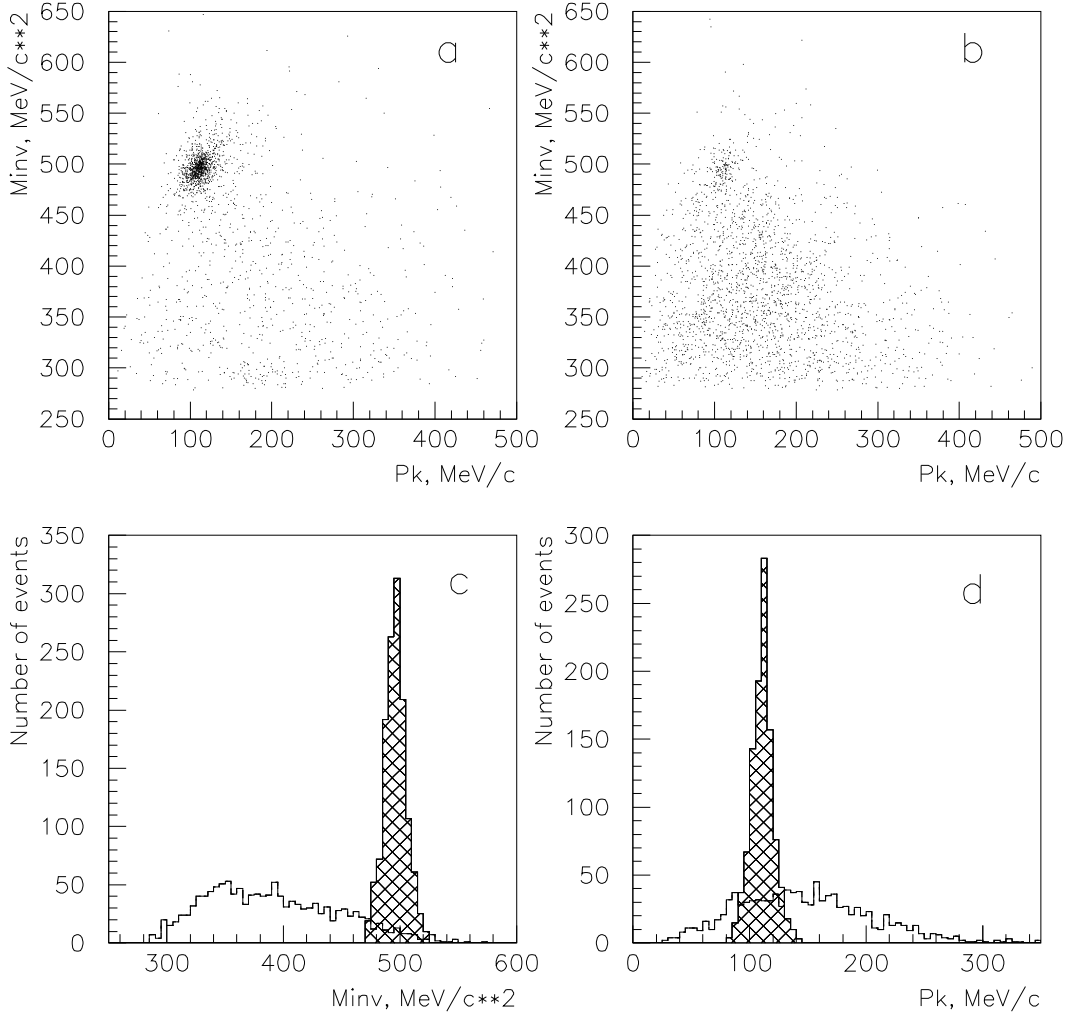,width=1.0\textwidth}}\\
\caption{Study of $K_{S}K_{L}$ coupled decays:
Invariant mass vs. missing momentum for 1-st (a) and 2-nd (b) vertex;
c. Invariant mass for $K_{S}$ (shaded) and $K_{L}$ after $K_{S}$ 
selection;
d. Missing momentum for $K_{S}$ (shaded) and $K_{L}$ after $K_{S}$
selection.}
\end{center}
\label{kskl1.fig}
\end{figure}

\hspace*{\parindent}
Candidates were selected from a sample in which two vertices,
each with two oppositely  charged tracks, were observed 
within 15 cm from the beam axes
and all tracks were reconstructed. An example of such an 
event is shown in Fig.~\ref{kskldec.fig}.
Figure~\ref{kskl1.fig}a shows a scatterplot 
of the invariant mass $M_{inv}$ of two charged tracks assuming
they are pions vs. the missing momentum $P_{mis}$ (which is equivalent to 
kaon momentum $P_{K}$ in case of two pion decay) 
for the particle which decays first. 
 Figure~\ref{kskl1.fig}b shows the same plot
for a particle from the second vertex.
The concentration corresponding to
$K_{S}$ mesons decaying into a pair of charged pions
dominates in the Fig. \ref{kskl1.fig}a and is seen in Fig. 
\ref{kskl1.fig}b showing events in which $K_{S}$ appears in the
second vertex.
The cuts  470 MeV $< M_{inv} <$ 525 MeV and 80 MeV/c $< P_{K} <$ 140 MeV/c
with an additional requirement to
have another reconstructed vertex in the $P_{mis}$ direction 
select events with $K_{S}$ in one of the vertices.  
The $K_{L}$ is expected to be in the other one. 
The above cuts represent three standard deviations
of the detector resolution.

Figures~\ref{kskl1.fig}c and \ref{kskl1.fig}d 
show peaked $M_{inv}$ and $P_{K}$ distributions corresponding to
$K_{S}$ after above cuts (shaded).
The $M_{inv}$ and $P_{K}$ parameters for the other vertex
give broad distributions 
expected for the main 3-body $K_{L}$ decays (into $\pi\mu\nu,~\pi e \nu,
~\pi^+\pi^-\pi^0$).

Figure~4a shows the decay length distribution 
for selected $K_{S}$. The exponential decay length is seen with
a correct value 0.55$\pm$0.02 cm and with a vertex position resolution 
of 0.15$\pm$0.03 cm.

The total number of the reconstructed $K_{L}$ vertices
accompanying $K_{S}$  decays was found to be 1423. 
The interpretation of the
decay radius distribution for the $K_{L}$ 
shown in Figure~4b
is more complicated.

The decay length for $K_{L}$ is about 350 cm, so only about 4\% of
$K_{S}K_{L}$'s both decay within 15 cm of DC and only half of them
both decay into charged modes. 
With the DC angular acceptance 
one can expect to detect 1\%  of $K_{S}K_{L}$ pairs if
the reconstruction efficiency is radially uniform.

But the reconstruction efficiency 
in the jet type DC cells
drops down for tracks with a vertex
radius greater than approximately 5 cm leaving only
0.64\% of all detected events with $K_{S}$ and $K_{L}$ vertices.
To describe the $K_{L}$ decay radius distribution, it is assumed that
the reconstruction efficiency is uniform to a certain radius and  
then drops down exponentially.

A clear peak at a radius of
1.7 cm corresponds to the $K_{L}$'s which
interacted with nuclei in the Be tube.
This peak has two contributions -  
$K_{L}$ regeneration into $K_{S}$ with
$K_{S}$ decay length seen afterwards and $K_{L}$ inelastic scattering
with two charged tracks seen in the final state.

 The combination of the described  processes was used to fit the 
experimental distribution and seems to describe it well enough. 
 The vertex spatial resolution obtained from the $K_{S}$ decay length
distribution was taken into account.
The radius at which the reconstruction efficiency 
begins dropping
down was found to be 5.0$\pm$0.4 cm with a 4.8$\pm$0.4 cm
visible exponential decay length.

%
 The number of $K_{L}$ having interactions in the Be tube was found to be
79 $\pm$ 18 with 1355 events representing $K_{L}$ decaying in flight. 

To select candidates to $K_{L} \rightarrow \pi^+\pi^-$ events,
an additional
cut requiring the invariant mass
of two tracks from a $K_{L}$ vertex to be in the range of 470-525 MeV 
was applied. 
The decay radius
%
%
distribution is presented in
Figure~4d together with the fit function 
where all parameters except
the number of events are fixed at the values obtained from the distribution in 
Figure~4b. 
The number of events under the peak drops down to 31 $\pm$ 7 
and 78 remain from decays in flight.
\vspace{-1.0cm}

The peak events after the invariant mass cut 
are interpreted
as regeneration of $K_{L}$ into $K_{S}$ with its decay into
$\pi^{+}\pi^{-}$ and are used for the calculation of the
regeneration cross section.

For the $K_{L} \rightarrow \pi^{+}\pi^{-}$ selection
one can apply stronger requirements for these events to satisfy
full kinematics within detector resolution, 
i.e. 80 MeV/c$< P_{K} <$ 140 MeV/c
and $K_{S}$ vertex in the $P_{mis}$ direction.
This selection is shown in Figure~4d by the shaded histogram and
demonstrates that the 
peak at the Be tube survives with 20$\pm$5 events and 35 $K_{L}$
decays in flight remain, still 10 times more than the expected
number of CP violating decays.

\vspace{-1.0cm}
\begin{figure}[H]
\begin{center}
\mbox{\psfig{figure=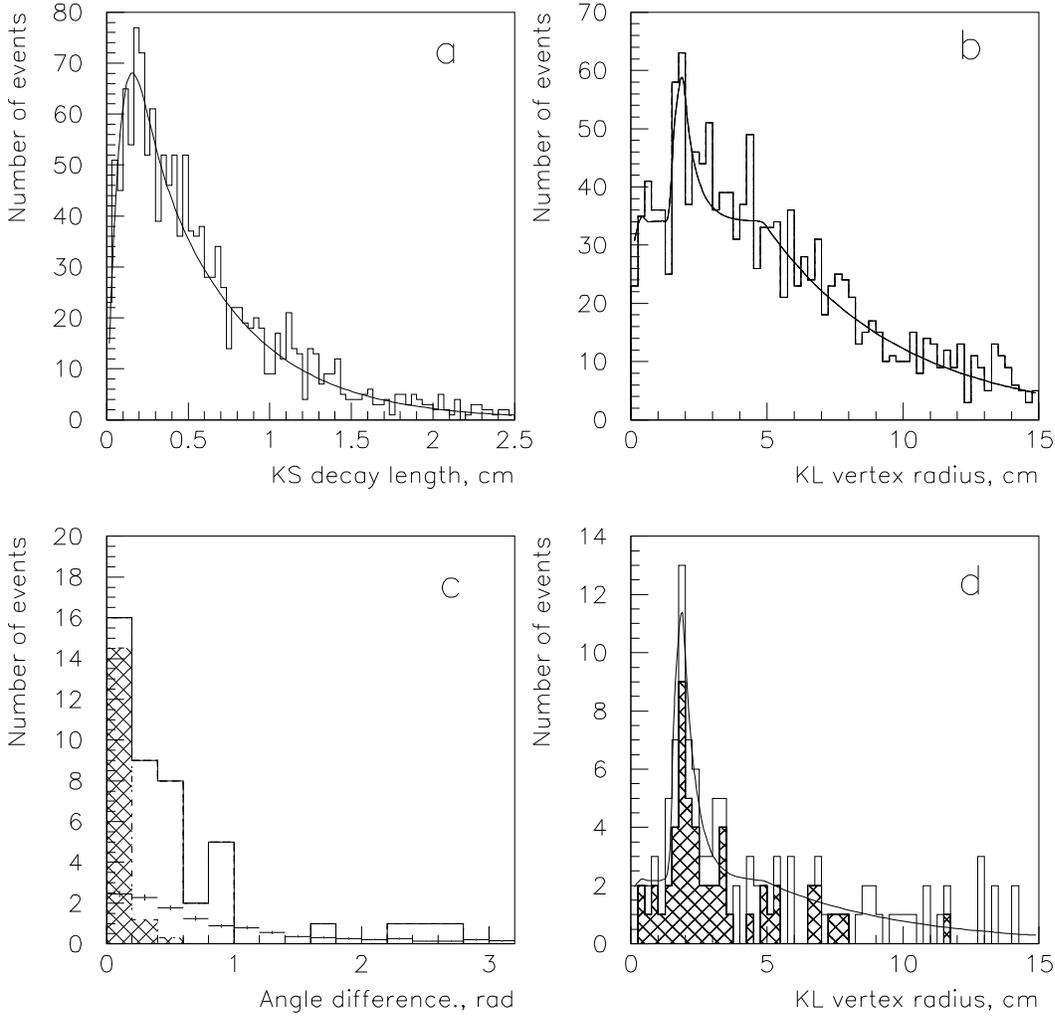,width=1.0\textwidth}}\\
\caption{Study of $K_{S}K_{L}$ coupled decays:
a. Decay length for $K_{S}$;
b. Decay radius for $K_{L}$;
c. Difference in the angle of $P_{mis}$ and vertex-beam line for 
"tube"events (histogram), $K_{L}$ semileptonic decays (points with 
errors and $K_{S}$ two pions decays (shaded); 
d. Decay radius for $K_{L}$s after $M_{inv}$ cut and after $K_{S}$ 
selecting cut (shaded).}
\end{center}
\label{kskl2.fig}
\end{figure}

The DC material (120 $\mu$ thick mylar entrance window, 100-150 $\mu$ diameter
50$\%$Cu-50$\%$Ti wires, Ar gas)
can also contribute to nuclear interaction events.
The DC mylar entrance window  
at 2.1 cm radius
adds about 10$\%$ to the tail of 
events peaked at the Be pipe. 
The biggest concentration of the DC field wires is between 2.5-3.5 cm,
where the average thickness of Cu+Ti is estimated as 
$3\times10^{-5} g/cm^{2}$ 
and drops down with radius. This amount of material is negligible
compared to Be and could produce less than one nuclear interacting 
event,
but a 10$\%$ correction from the mylar window is applied to cross section. 

A further correction to the $K_L$ cross sections comes from the 
$\Lambda$ and $\Sigma$ hyperon decays which can appear in the
kinematic range as $K_L \rightarrow 2\pi$. The reactions
\\
\\
$K_{L} + p \rightarrow \Lambda^{0} + \pi^{+},  
\Lambda^{0} \rightarrow p + \pi^{-}$;
\\
$K_{L} + p \rightarrow \Sigma^{0} + \pi^{+},  
\Sigma^{0} \rightarrow \Lambda^{0} + \gamma, \Lambda^{0} \rightarrow p + \pi^{-}$;
\\
$K_{L} + n \rightarrow \Sigma^{-} + \pi^{+},  
\Sigma^{-} \rightarrow n + \pi^{-}$;
\\
$K_{L} + n \rightarrow \Sigma^{+} + \pi^{-},  
\Sigma^{+} \rightarrow n + \pi^{+}$
\\
\\
were studied. In our selection only two charged pions were detected 
by the requirement of two prong vertex and dE/dX in the DC corresponding to
minimum ionizing particles. 
 
As it was shown by simulation, 8$\%$ of such events 
with a final nucleon nearly at rest
contribute to
the number of regenerated events 
after the $M_{inv}$ cut with the visible vertex radius distribution similar
to that from the $K_S$ decay.
These events were removed from the regeneration candidates in
the $K_L$ kinematic region and added to 79 - 31 = 48 
$K_L$ interaction candidates.
 
The following numbers were obtained:
\\
\\
$N_{nucl+reg} = 79 \pm 18$,\\
$N_{nucl} = 50 \pm 17$,\\
$N_{reg}  = 29 \pm 7$.\\
\\

Events in the range 1.5-3.5 cm from the distribution in Figure~4d were used
to obtain the angular distribution for $K_{S}$ after regeneration.
Figure~4c shows the 
projected angular distribution of these 
$K_{S}$.
Dots with errors show the expected distribution for
semileptonic decays of $K_{L}$, normalized to the expected number of
these background events in the selected range. 

The obtained angular distribution is wider than
in the case of coherent regeneration which should look like shown
shaded
distribution  for original $K_{S}$ decaying at the same distance. 
But with this data sample the coherent contamination 
is expected to be small and
cannot be extracted.
\section{ Cross sections }
\hspace{0.5cm}For the calculation of the cross
sections, the number of the initial $K_{L}$ passing the Be tube
multiplied by the reconstruction efficiency $\epsilon_{rec}$ 
can be found from the distribution of Figure~4b using the expression:
\\
\\
$N_{K_{L}} \cdot \epsilon_{rec} =  N_{0} \cdot L_{K_{L}}/B_{ch} = 61200 \pm 2500$,
\\
\\
where  $N_{0}$ 
is the visible number of pairs at a zero radius 
(fit value from the flat region in Figure~4b)
and $L_{K_{L}}$=350 cm is the
decay length for the $K_{L}$.
The value $B_{ch}$=0.78 
is the probability for a $K_{L}$ to have a
pair of charged particles after decay.
Using this expression one can
calculate a probability for the $K_{L}$ to interact at the Be tube as\\
\\
$P_{nucl} = N_{nucl}/(N_{K_{L}} \cdot \epsilon_{rec}) = (0.82 \pm 0.28) \cdot 10^{-3}$,
\\
$P_{reg} = N_{reg}/(N_{K_{L}} \cdot \epsilon_{rec} \cdot B_{\pi\pi}) = (0.61 \pm 0.16) \cdot 10^{-3}$,
\\
\\
where $B_{\pi\pi}$=0.686 is a branching ratio of $K_{S}$ decays 
into a pair of charged pions. 
A ratio $(1.12 \pm 0.16)$ for the reconstruction 
efficiencies of $K_{L}$ decaying into
two pions (after regeneration) and into three particles
was obtained by simulation and was taken into account.

Using the above probabilities the cross sections can be calculated 
from the equation\\
\\
$\sigma = \frac {P \cdot A}{N_{A} \cdot \rho \cdot t}$ ,
\\ 
\\
where A and $N_{A}$ are
the atomic number and the Avogadro constant, $\rho$ and t are 
the density and the thickness of the material. 
After correction for the mylar window interactions
the following cross section for Be has been obtained:\\
\\
$\sigma_{reg}^{Be}$ = 53 $\pm$ 17 mb.
\\
\\
For the nuclear cross section 
(excluding regeneration)
one can obtain the inelastic
cross section into a two particle final state,
arising from the $\Lambda$ and $\Sigma$ production discussed earlier:
\\
\\
$\sigma_{inel}^{vis}$ = 70 $\pm$ 26 mb.
\\
\\

To estimate the total cross section, the relative
weight of these reactions was found  to be 0.21
from the NUCRIN package \cite{nucrin}. 
Using the CMD-2 experimental $\sigma_{inel}^{vis}$ 
value 
and a ratio $\sigma_{inel}/\sigma_{tot}$ = 0.52 \cite{andrea},
one can estimate  $\sigma_{tot}^{Be}$ = 641 $\pm$ 238 mb. A systematic 
error of this estimation is about 30$\%$ and comes from uncertainties in
the event selection and in the ratio above.
\section{ Discussion}
\hspace{0.5cm}
The selection of candidates for $K_{L}\rightarrow\pi^+\pi^-$ events 
faced two problems. 

First is a background from the dominant $K_{L}$ decays.
Simulation gives a rejection factor of only 0.03 for the given DC
resolution, and the ratio of the number of events within our cuts
(away from the beam pipe) to the total number of $K_{L}$ decays,
35/1355, is consistent with this result. However, the expected rate of
$K_L \rightarrow 2\pi$ decays is $ 2 \cdot 10^{-3}$, so that 
only 3 true $K_L \rightarrow 2\pi$ events could be expected in 
the sample.
The observed events are mainly from the
semileptonic decays of $K_{L} \rightarrow \pi\mu\nu$, and
to  improve the achieved rejection and select 
$K_L \rightarrow \pi^+\pi^-$ events
better angular and momentum resolution in the DC is needed. 

The effect of tightening resolution
may be illustrated by
selecting events within
1.5 standard deviations of the
detector resolution.

In this case 9.5$\pm$3.4 
of the original
events remain under the peak at the pipe
and  only 5 $K_{L}$ decays in flight remain 
(2 of them should be real $K_{L} \rightarrow \pi^{+}\pi^{-}$ 
events). 
DC resolution better by a factor of 2 expected for the rest of 
our data, could give a signal/background ratio about 1 or 
better.

A second problem is the relatively high background from nuclear interactions
of $K_{L}$ and the regeneration effect.
The obtained cross sections can be compared 
to the data available at
higher momenta and with theoretical expectations. There are 
no data for the regeneration cross section for Be and for slow kaons. 

\begin{figure}[tbh]
\begin{center}
\mbox{\psfig{figure=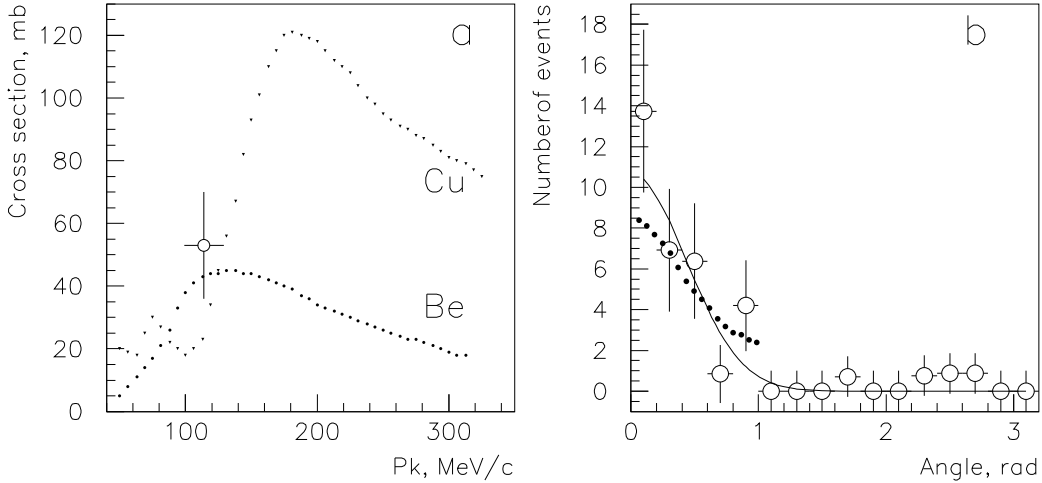,width=1.0\textwidth}}\\
\vspace{-7.5cm}
\caption{
a. Experimental regeneration cross section and theoretical calculations
for Be and Cu;
b. Angular distribution of the regenerated $K_{S}$ with the best fit
function (solid line) and theoretical prediction (dots);}
\end{center}
\label{kskl3.fig}
\end{figure}

 In Figure~\ref{kskl3.fig}a the experimental 
regeneration cross section is plotted together
with the theoretical calculations performed in \cite{andrea}. The obtained
cross section is consistent with calculations. The calculated
regeneration cross section for Cu is also presented in the Figure. The 
comparison of the calculated regeneration cross sections for these two 
different materials shows
that at momenta below 200 MeV/c 
one cannot scale them by a simple $A^{2/3}$ dependence.
 To verify these theoretical calculations one needs experimental measurements
for different materials.
%
 The experimental angular distribution of the regenerated $K_{S}$ after
background subtraction is presented in Figure~\ref{kskl3.fig}b together
with the fit function and the theoretical prediction \cite{andrea}
and seems to be a little narrower than the expectation.

 This result for the total nuclear cross section in Be
for  110 MeV/c kaons can be compared 
to the experimental data at 
higher momenta \cite{bedata} and theoretical calculations 
\cite{andrea}. Such comparison is presented in 
Figure~6. The cross sections extracted from GHEISHA 
and FLUKA simulation codes are 
also shown. It is seen that the FLUKA code, as well as the calculations
from \cite{andrea} give cross sections in good agreement with 
experimental data. 
The GHEISHA code gives completely wrong absolute values as well as 
momentum dependence.

After publishing our preliminary results \cite{CMD2pre93}, 
the regeneration influence was discussed for the $\epsilon'/\epsilon$
measurement planned in KLOE detector\cite{andrea}. 
A possible way to remove regeneration events is to require that 
the final state particles lie in a plane (since for regeneration
events, recoil nucleons will carry a momentum).
It was shown that 
the total regeneration probability in the KLOE drift chamber 
after an acoplanarity cut (factor of 4 rejection) was $10^{-4}$ that
should be compared with $2\times10^{-3}$ probability for
the "normal" CP violating
$K_{L}\rightarrow\pi\pi$  decay and $\approx10^{-6}$ probability for
the expected direct CP violation decay. 

The regeneration itself does not give any decay 
asymmetry expected for the direct CP violating $K_{L}$ decay, but
the acoplanarity cut as well as other selection cuts applied separately
to $\pi^+\pi^-$ and $\pi^0\pi^0$ final states with a different
resolution (DC for the first and calorimeter for the second) can cause
a systematic asymmetry due to the broad angular distribution of the
regenerated events.
%
\begin{figure}[tbh]
\begin{center}
\mbox{\psfig{figure=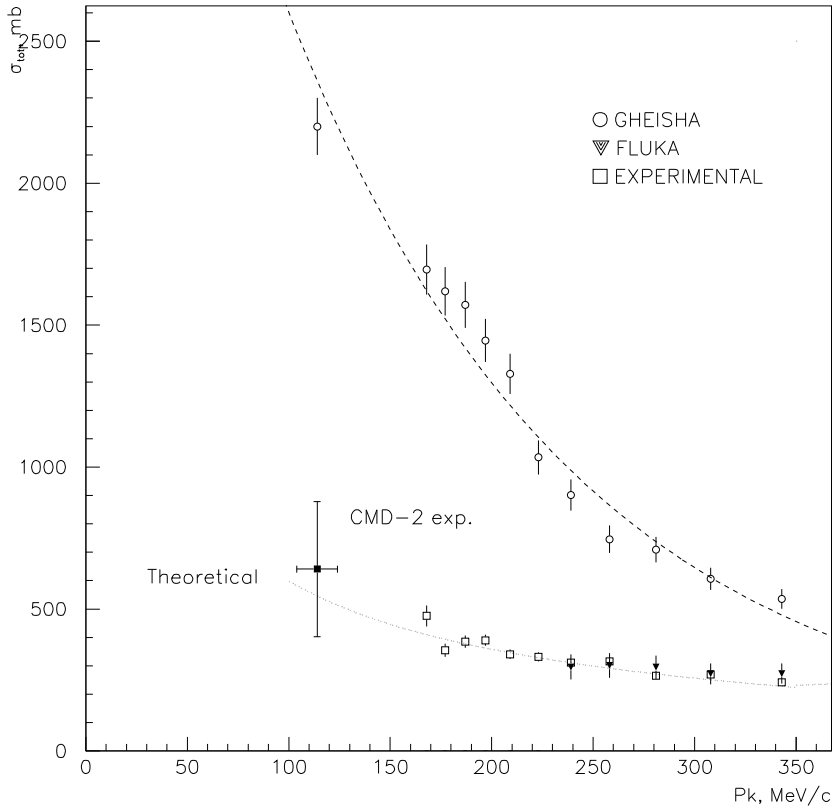,width=0.8\textwidth}}\\
\caption{
Comparison of the total experimental $K_{L}$ nuclear interaction 
cross section in Be with
the theoretical calculations and simulation by different codes.
}
\end{center}
\label{kskl4.fig}
\end{figure}
\section{Conclusions}
\hspace*{\parindent}

With the CMD-2 detector the coupled $K_{S}K_{L}$
decays from the $\phi$ have been observed for the first time
at an $e^+e^-$ machine.

The attempt to select
$K_{L} \rightarrow  \pi^{+}\pi^{-}$ events 
emphasized  again
problems of the $K_{L}\rightarrow\pi\mu\nu$
decay mode background
as well as a high
level of nuclear interactions of neutral kaons including 
$K_{L}$ into $K_{S}$ regeneration.

The regeneration and total nuclear interaction cross sections for low momenta
neutral kaons have been measured.
The measured values of the cross 
sections indicate that regeneration 
will cause an additional
background  
for the CP-violating decays of $K_{L}$ at 
the $\phi$-factory experiments and should be 
carefully studied to avoid systematic errors. 

Comparison of the obtained total cross section  with the predictions of 
the simulation codes GHEISHA, FLUKA and calculations performed in 
\cite{andrea} 
shows that GHEISHA results are
wrong while two other codes are in good agreement with experimental 
data. 

The opening of this kinematic region for studies of
the neutral and charged kaon interactions has been an
additional argument for the construction of $\phi$-factories and
the detector FINUDA \cite{finuda} has been proposed for 
these studies at the $\phi$-factory in Frascati\cite{vignola92}. 
We anticipate
that the  results obtained from the data now in hand will be 
important in planning of experiments at $\phi$-factories.

The analysis of new experimental data
is in progress and we expect new results on the  nuclear interaction cross 
section.
\section{Acknowledgements}
\hspace*{\parindent}
One of the authors (E.P.Solodov) would like to thank R. Baldini 
for useful discussions on the nuclear interaction studies and cooperation
in this field and A. Michetti for simulation of the hyperon production
of neutral kaons and theoretical cross section calculations. 

This work is supported in part by the US 
Departament of Energy, US
National Science Foundation and the
International Science Foundation under grants RPT000 and RPT300.

\begin{thebibliography}{99}

\bibitem {bayer73}{V.N.Bayer, ZETFP {\bf{17}} (1973) 446.}

\bibitem {CMD285}{G.A.Aksenov et al., Preprint
BudkerINP 85-118, Novosibirsk, 1985.}

\bibitem{vepp2m} {V.V.~Anashin et al., Preprint BudkerINP, 84-114, 
Novosibirsk 1984.}

\bibitem{epr}
{ P. Eberhard, Contribution to the  $\phi$ Factory Workshop at UCLA,
     April, 1990.;             
     G.Ghirardi,  Proceedings of the Workshop on Physics and Detectors
     for DA$\Phi$NE, Frascati, April, 1991, p.261.}

\bibitem {rosner}{J.L. Rosner, I. Dunietz, J. Hauser, Phys.
Rev. {\bf{D35}} (1987) 2166.}

\bibitem {phigroup}{A partial list of other major contributions includes:
A.N.Skrinsky {\it{et al}}., "Novosibirsk $\phi$-factory project",
F.J.  Botella, J.  Bernabeu  and  J.  Roldan -Blois  CP
Violation Conference, FTUV/89-35,IFIC/89-11, May 1989.;
G. Barbiellini and C. Santoni, CERN-EP/89-8 and CERN-PPE/90-124;
J.A. Thompson, University of Pittsburgh preprint PITT-90-09;
     contributions by J.A. Thompson, D. Cline, and by C. Buchanan and others
      to the $\phi$ Factory Workshop at UCLA, April, 1990.;
     Y. Fukushima, {\it{et al}},.KEK Preprint 89-159;
      summary talks by P. Franzini and M. Piccolo and others at the
Workshop on Physics and Detectors for DA$\Phi$NE, Frascati, April, 1991.}

\bibitem {skrinsky91}{A.N.Skrinsky, Proceedings of the Workshop on Physics
and Detectors for DA$\Phi$NE, Frascati, April, 1991, p. 67.}

\bibitem {vignola92}{G. Vignola, Proceedings of the Workshop on Physics
and Detectors for DA$\Phi$NE, Frascati, April, 1991, p. 11.}

\bibitem{round}{A.N.Filippov et al., Proceedings of the XVth
International Conference on High Energy Accelerators,
Hamburg, Germany, 1991, World Scientific, V. II, p. 1145.}

\bibitem{est90}{ S. Eidelman, E. Solodov, and J. Thompson,
        Nuclear Physics B (Proceedings Supplement) {\bf 24A} (1991) 174.}

\bibitem {cmd2gen}{E.V. Anashkin et al.,
     ICFA Instrumentation Bulletin 5 (1988), p.18.}

\bibitem {PhLet95}{R.R. Akhmetshin et al., Phys. Lett.{\bf B 
364}(1995) 199.}

\bibitem  {CMD2pre93}{R.R. Akhmetshin et al., Preprint BudkerINP 95-62,
 Novosibirsk, 1995.}

\bibitem{pdg}{R.M.Barnett et al., Phys.Rev.{\bf D54}(1996)1.}

\bibitem {andrea}{R. Baldini, A. Michetti, Preprint LNF-96/008, 1996.}

\bibitem {nucrin}{This value was extracted from CERN GEANT package.}

\bibitem {bedata} {G. A. Sayer and E. F. Beall, Phys. Rev. 
{\bf 169}(1968)1045.}

\bibitem {finuda}{M.Agnello et al. (FINUDA collaboration),
     FINUDA, proposal LNF-93/021 (IR) 11 Maggio 1993.}

\end{thebibliography}
\end{document}